# All-solid flexible fiber-shaped lithium ion batteries

*Hang Qu[1,∆], Xin Lu[2,∆], and Maksim Skorobogatiy*[1,2]*.

[1]Génie Physique and [2]Génie Métallurgique, École Polytechnique de Montréal, Montreal, QC, Canada, H3T 1J4. [∆]Contributed equally to this work.

ABSTRACT: We propose fabrication of the fiber-shaped lithium ion batteries assembled by twisting a cathode filament together with an anode filament. The cathode filament is fabricated by depositing a LiFePO$_4$ (LFP)-composite layer onto a steel-filled polyester conductive thread (SPCT). As anode filaments, we propose several scenarios including a Li$_4$Ti$_5$O$_{12}$ (LTO)-composite coated SPCT (dip-and-dry deposition), a tin-coated SPCT (PVD deposition) as well as a bare tin wire. An electrolyte composite layer consisting of LiPF$_6$ and polyethylene oxide (PEO) is then deposited onto both the anode and cathode filament before the battery assembly. By twisting the cathode filament and anode filament together using a customized jig, the batteries are then assembled. The open-circuit voltage is found to be ~ 2.3 V for the battery using the LTO@SPCT anode, and ~3.3 V for the battery using the tin@SPCT anode and the tin wire anode. Charge-discharge tests are carried out at different C rates for each battery sample. Experimental results suggest that the LIBs using the LTO@SPCT anode, the tin@SPCT anode and the bare tin wire anode could achieve a specific capacity of ~64, ~67, and ~96 mAh/g, respectively, when charge-discharged at 0.5-C rate. The battery could retain well its capacity after 80 charge-discharge cycles. During operation of all the batteries reported in this paper, their coulombic efficiency remained above 80%. The reported batteries were also tested mechanically and we found that their electrochemical performance did not show signs of degradation even after ~30000 bend-release cycles. The fabrication of the proposed LIBs is simple and cost-effective, as compared to the fiber-shaped LIBs using carbon nanotube fibers. Moreover, the reported LIBs are well suitable for the wearable applications as they feature all-solid electrodes and electrolyte, unlike the majority of other currently existing LIBs that utilize liquid organic solution-based electrolytes that may cause leakage and cause safety concerns. Among other advantages of the proposed LIB are light weight, ease of fabrication, high specific capacitance, high energy density, and good durability. Finally, employing cheap and commercially-available steel-filled polyester threads as a base material in our batteries, makes them potentially suitable for integration into wearables using various standard textile manufacturing techniques.

**Introduction**

Most recently, flexible, portable and wearable electronic devices have received significant attention in a view of their potential applications in several emerging markets [1-3]. Many conceptually new products such as on-garment displays, wearable sensors for sports and medicine, virtual-reality devices, smart phones, smart watches and bracelets recently became commercially available, and are utilized by an ever-growing number of people. Currently, most of these personal wearable electronic devices use conventional lithium-ion batteries (LIBs) as power sources. Although these LIBs have many advantages such as high energy density, high output voltage, long-term stability and environmentally friendly operation [4-5], they are heavy and rigid, and thus not truly compatible with wearable applications. Therefore, the R&D of flexible and wearable LIBs in the form of fibers or ribbons is in rapid expansion, as slender flexible LIBs could be weaved seamlessly into textiles or fabrics for wearable applications, or alternatively, packed into compact power source units.

Many attempts have been made to fabricate fiber- (wire-) shaped LIBs [6-16]. For example, Neudecker *et al*. proposed a fiber LIB which was fabricated by sequential deposition of battery-component thin-layers (such as in sequence of anode, electrolyte, cathode layer) on a conductive fiber substrate [6]. They also embedded multiple fibers into an adhesive matrix to constitute a battery ribbon. Note that the fabrication of these fiber batteries requires complicated material deposition techniques such as magnetron sputtering and electron-beam evaporation which are inevitably operated in high vacuum environment. Wang *et al*. reported fiber-shaped LIBs fabricated using intrinsically conducting polymers [7]. To fabricate the batteries, they electropolymerized polypyrrole-hexafluorophosphate (PPy/PF$_6$) on a platinum (Pt) wire as the cathode. Then, the cathode filament was inserted into a hollow-core polyvinylidene fluoride (PVDF) membrane separator winded by a polypyrrole-polystyrenesulfonate (PPy-PSS)-coated Pt filament as the anode. Finally, the whole structure was immersed in a glass vial filled with an electrolyte of lithium hexafluorophosphate (LiPF$_6$) solution. The battery capacities on the order of 10 mAh/g over 30 cycles were reported. Later, they replaced PPy-PPS by single-walled carbon nanotubes (CNT) for the fabrication of anode [8]. The capacities were thus improved to ~ 20 mAh/g. Both of these fiber batteries, however, should be immersed in liquid electrolytes to function, which makes them unsuitable for wearable applications.

Recently, Kwon et al. reported a cable-type flexible LIB that consists of a hollow spiral, spring like anode (comprising nickel-tin coated copper wires), a lithium cobalt oxide (LiCoO$_2$) cathode, and a poly(ethylene terephthalate) (PET) nonwoven separator membrane [9]. After encapsulating the electrode filaments into a heat-shrinking tube, 1M LiPF$_6$ solution was injected into the battery as electrolyte. This cable-type battery has a linear capacity of ~ 1 mAhcm$^{-1}$. H. Peng and coworkers proposed several flexible, stretchable LIBs that are fabricated by parallel winding the anode filament and cathode filament into a spring-like structure around an elastic fiber substrate [10, 13, 16]. The anode and cathode filaments were fabricated using CNT/lithium titanium oxide (LTO) composite and CNT/lithium manganate (LMO) composite, respectively. A gel electrolyte comprising lithium bis(trifluoromethane)sulfonamide (LiTFSI)/polymer composite was coated onto the two electrode filaments, and this gel electrolyte also functioned as a battery separator. Thus fabricated stretchable LIBs could retain well their electrochemical performance even after several hundreds of stretch-release cycles, and their capacity was up to ~90 mAh/g. Quite recently, the same group also demonstrated a fiber-shaped aqueous lithium ion battery that used a polyimide/CNT fiber as the anode and a LMO/CNT fiber as the cathode [15]. The anode and cathode fibers are encapsulated into a heat-shrinking tube and then an aqueous $Li_2SO_4$ solution was injected into the tube as an electrolyte. Thus fabricated LIB featured a power density of ~10000 W kg$^{-1}$. Note that the use of aqueous electrolyte solution could avoid safety issue caused by the flammable organic electrolytes; however, the liquid electrolyte may leak out of the battery, thus leading to pollution and degradation of the battery performance.

On a general note, we remark that many of the high-performance fiber-shaped LIBs currently utilize CNT fibers (yarns) in the electrode fabrication. Thought CNT fibers feature outstanding electrical and mechanical properties, the production of CNT fibers is extremely expensive, and often require using expensive high-vacuum deposition instruments and techniques. Thus, the search for alternative cheaper conductive threads for the LIB electrodes is critical, especially for high-production-volume wearable electronics applications.

Moreover, most of the current high-performance LIBs use liquid organic electrolytes and require careful handling during operation as electrolyte leakage could cause pollution and pose health risks to the users. Therefore, the development of LIBs featuring gel [11, 13, 16] or solid electrolytes constitute an important research field in wearable and compliant batteries.

In this paper, we report fabrication of the high-performance, mechanically and electrochemically stable fiber batteries for wearable applications. Our batteries use commercially available and cost-effective conductive yarns as well as all-solid electrolytes. The flexible fiber-shaped LIBs are assembled by co-twisting a cathode filament and an anode filament. The cathode filament is fabricated by depositing a $LiFePO_4$-PVDF composite layer onto a steel-filled polyester conductive thread (cathode abbreviated for LFP@SPCT) using the dip-and-dry method. As to the cathode filament, we have tried successfully three different material combinations that include: a tin-coated SPCT (abbreviated as tin@SPCT), a LTO-PVDF composite-coated SPCT (abbreviated as LTO@SPCT), and a bare tin wire, respectively. Furthermore, an all-solid $LiPF_6$-polyethylene oxide (PEO) composite layer that functions as both the electrolyte and battery separator is deposited onto cathode and anode filaments before the battery assembly. Titanium oxide ($TiO_2$) nanoparticles are doped into the electrolyte layer in order to lower the polymer crystallinity and increase the ionic conductivity [17]. The electrochemical performance of the proposed fiber LIBs is characterized using standard C-rate charge-discharge tests. Experimental results show that the LIBs using the LTO@SPCT anode, the tin@SPCT anode and the bare tin wire anode achieve specific capacities of ~64, ~67, and ~96 mAh/g, respectively, when charge-discharged at 0.5-C rate. Additionally, our batteries retain their capacity after 80 charge-discharge cycles. During operation of all our batteries, their coulombic efficiency always remained above ~80%. Moreover, the fabricated batteries also went through a series of bending tests that included high amplitude bend-release motions of the fiber batteries, while the batteries were cyclically charge-discharged at 1 C rate. Experimentally, the batteries maintained well their electrochemical properties even after ~30000 bend-release cycles. Finally, we note that fabrication of the reported LIBs is cost-effective and simple as it involves readily-available commercial materials as well as simple material processing. Our LIBs feature all-solid electrodes and electrolytes thus avoiding safety concerns associated with electrolyte leakage. Among other advantages of the reported LIBs are light weight, good flexibility, high specific capacitance, high energy density, and good durability.

**Fabrication and characterization of the fiber-shaped LIBs**

(1) **LIBs comprising a LFP@SPCT cathode and a LTO@SPCT anode**

We now detail fabrication of the LIB using a LFP@SPCT cathode and a LTO@SPCT anode. To fabricate the LIB cathode filament, a LFP-PVDF composite solution was first prepared by dissolving LFP and PVDF in 1-Methyl-2-pyrrolidinone (NMP) solvent. Carbon-nanofiber powders were added into the cathode solution to increase the cathode electric conductivity. Then, the LFP composite layer was deposited onto a SPCT using dip-and-dry method (Fig. 1(a)). Similarly, to fabricate the LTO@SPCT anode, we first prepared a LTO-PVDF solution by dissolving LTO, PVDF, and carbon nanofiber powders in NMP solvent. Then, a LTO composite layer was deposited onto a SPCT also using dip-and-dry method. Before the battery assembly, a $LiPF_6$-PEO composite electrolyte layer that also functions as the battery separator was coated on the anode and cathode filaments using dip-and-dry method (Fig. 1(b)). The electrolyte solution was prepared by dissolving $LiPF_6$ and PEO in acetonitrile solvent. Additionally, $TiO_2$ was also added into the electrolyte solution in order to lower the polymer crystallinity and improve the electrolyte ionic conductivity.

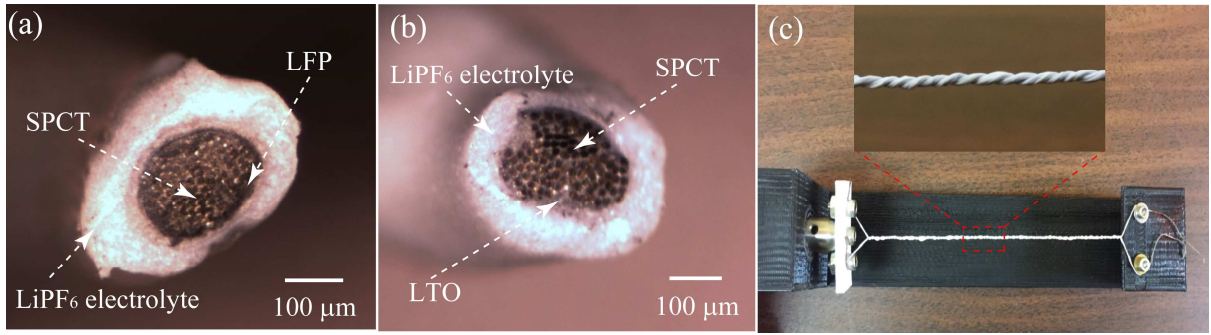

**Figure 1**. (a) Cross section of the LIB cathode filament; (b) Cross section of the LIB anode filament; (c) Two electrode filaments are co-twisted using a home-made jig. Insert: an enlarged view of the twisted filaments.

As shown in Fig. 1(c), the two electrode filaments are co-twisted in a controllable fashion using a home-made jig. In order to enhance bonding between the two electrolyte layers, we also wetted the fabricated battery with several drops of propylene carbonate. Finally, the battery was encapsulated within a heat-shrinking tube at 120 ℃ for 60 seconds. Electrode preparation, as well as the battery assembly were all carried out inside the N$_2$-filled glove box.

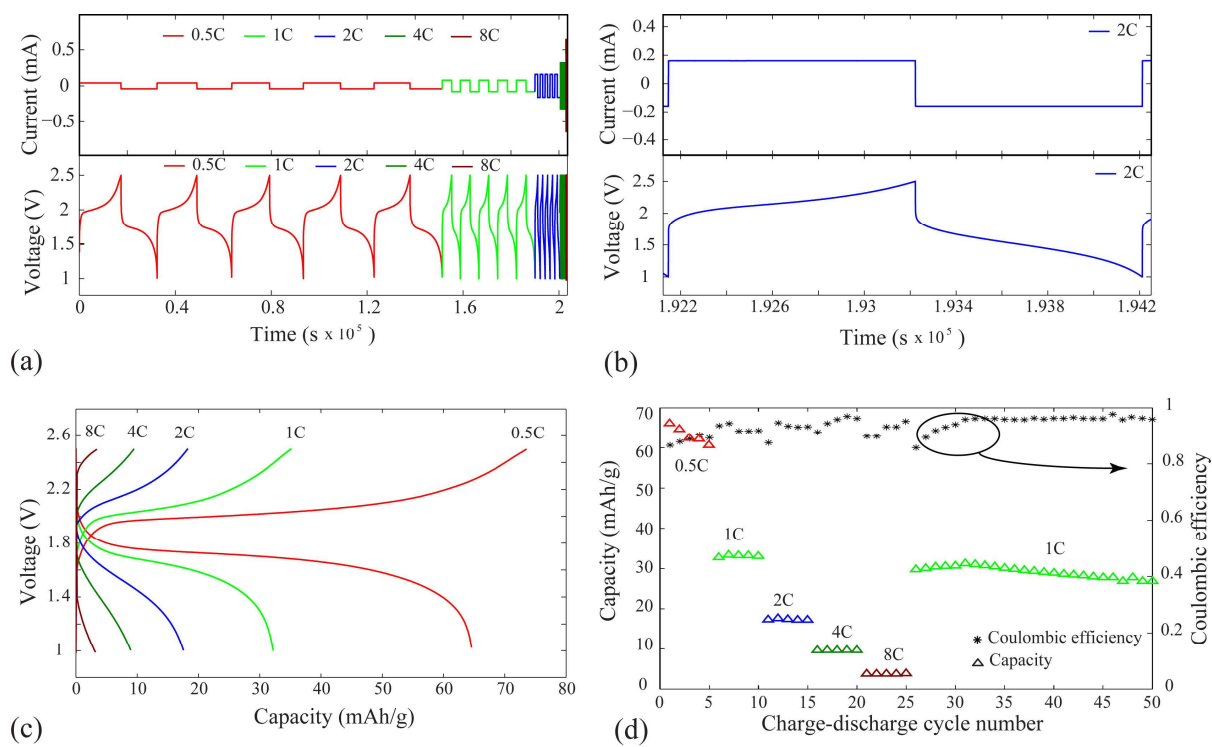

**Figure 2**. Electrochemical performance of the battery using LTO@SPCT anode. (a) Cell voltages and currents in the first 25 cyclic charge-discharge tests with different C-rates. (b) A zoomed view of the cell voltage and current in a charge-discharge test with a 2 C rate. (c) Battery voltages in different charge-discharge cycles as a function of the battery capacity. (d) Specific capacity and coulombic efficiency of the battery measured in 50 charge-discharge cycles with different C rates.

The electrochemical properties of a full fiber-shaped LIB using the LTO@SPCT anode were investigated using a cyclic charge-discharge analysis with different charge-discharge rates (from 0.5 C to 8 C) as shown in Fig. 2 (a, b). The measurements were conducted using an Ivium Electrochemical Workstation and ~12 cm-long batteries stored in the $N_2$-filled glove box. The battery open-circuit voltage was ~ 2.1 V. As seen in Fig. 2 (c, d), the battery features capacities of 64.1, 33.7, 17.7, 8.5, 3.8 mAh/g at 0.5, 1, 2, 4, and 8 C charge-discharge rates respectively. We note that the battery capacities decrease rapidly with increased current rates (Peukert constant of ~2). This is because the practical ionic current density of the electrolyte and electrodes, including the rate of ion transfer across the electrode/electrolyte interface, is generally much smaller than the electronic current density of the external electronic circuit [18]. Therefore, at a high charge/discharge current rate, the ionic motion within an electrode and/or across an electrode/electrolyte interface is too slow for the charge distribution to reach equilibrium, thus leading to a decreased capacity. As the charge-discharge current rate decreased, the capacity loss was also recovered. Besides, we like to mention that this LIB still has the voltage plateau that remained flat even at a current rate as high as 8C, indicating good charge-discharge performance. After a series of tests with different C rates between 0.5 C and 8 C, we performed an additional charge-discharge experiment using 1 C as a discharge rate and noted that the capacity of the LIB returned to the expected ~ 29.3 mAh/g. This indicates that the battery internal electrochemical structure remained intact even after been subjected to high current rates. Finally, 25 cycles of the charge-discharge tests at 1 C rate were performed (see Fig. 2 (d)), and we noted that the battery still retained 83% of its original capacity after these cyclic charge-discharge tests. The coulombic efficiencies of the battery were greater than 83% during all the charge-discharge cycles, while mostly staying above ~93% (see Fig. 2 (d)).

The Electrochemical Impedance Spectroscopy (EIS) studies of the battery were then performed. The Nyquist plot of a battery (see Fig. 3 (a)) is composed of a depressed semicircle in the high-to-medium frequency region followed with a slope in the low frequency region. This is a classic shape of the EIS that can be fitted using an effective electric model shown in the insert of Fig. 3(a). There, $R$ denotes various Ohmic resistances, while $W$ denotes Warburg impedance. The first intercept of the EIS curve with the real axis in Fig. 3 (a) gives the equivalent series resistance $R_s$ ~208 Ω, which is a bulk electrolyte (ionic) resistance. The second intercept gives a sum of the electrolyte resistance $R_s$ and the charge transfer resistance $R_{ct}$, which is the electrode-electrolyte interfacial resistance. From Fig. 3 (a) we find that $R_{ct}$~856 Ω. We note that our wire-shaped batteries featuring the all-solid structure have a linear resistivity of ~ 120 Ω·m that is higher than resistivity (5-30 Ω·m) of the fiber batteries using aqueous electrolytes [9, 15].

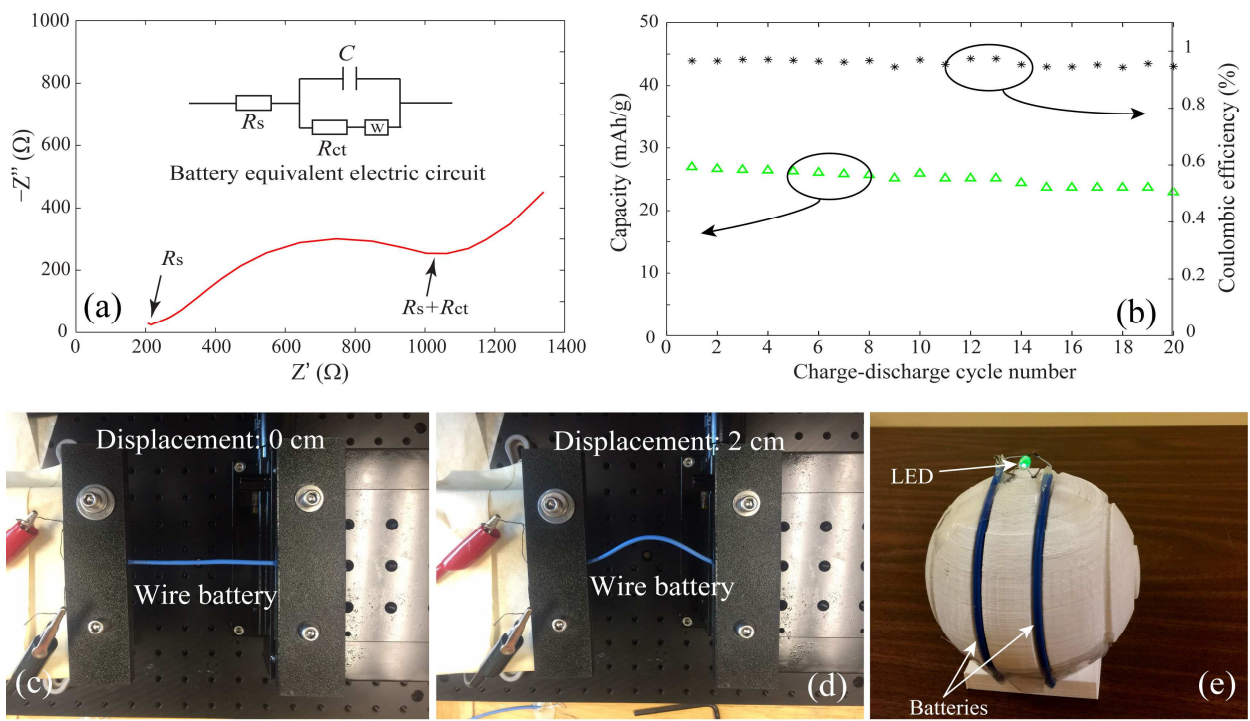

**Figure 3**. (a) EIS spectrum of the battery using an LTO@SPCT anode. Insert: the equivalent electric circuit of the battery. (b) Battery specific capacity and coulombic efficiency during bending test. (c-d) Experimental setup to perform bending tests. In a single bend-release movement, one end of the battery is fixed, while the other end is displaced by 2 cm and then returned to its original position. In total, 31000 bend-release movements are carried out during 20 charge-discharge cycles with 1 C current rate. (e) Two batteries using the LTO@SPCT anode are used to light up an LED. The batteries are immobilized in the grooves of a sphere fabricated via 3D printing.

Finally, we tested resilience of the battery electrochemical properties to mechanical influence by performing a large number of bend-release movements. Experiments were carried out outside of the glove box using a LIB packaged inside of a shrink tube. The 20 charge-discharge cycles of the battery were performed at 1 C rate, while the battery was simultaneously subjected to 31000 bend-release movements. In a single bend-release movement, one end of a ~12cm-long battery is fixed while the other end is displaced by 2 cm and then returned to its original position (Fig. 3(c, d)). The period of a single bend-release movement was 4 s, while a typical battery charging-discharge cycle period was ~103 min. As seen from Fig. 3 (b), after 31000 bend-release movements, the specific capacity of the battery decreased only by ~14% while the battery coulombic efficiency remained practically constant ~93% during the whole experiment. In Fig. 3(e), we also demonstrate to use two batteries to light up an LED. Thanks to their flexibility, the two batteries could be easily immobilized in the grooves of a sphere fabricated via 3D printing.

### (2) LIBs comprising a LFP@SPCT cathode and a tin@SPCT anode

As an alternative scenario for the fabrication of fiber-shaped LIBs, we replace the LTO@SPCT anode with a tin@SPCT anode. The anode was fabricated by coating a SPCT with a thin layer of tin using the Physical Vapour Deposition (PVD) technique. Particularly, Edwards Inc. evaporator was used to deposit a ~1.6 μm thick tin layer on the surface of a conductive thread. LFP@SPCT cathode and electrolyte were fabricated as described in the previous section; the cathode and anode filaments were coated with the electrolyte layers via a dip-and-dry process and assembled into a battery inside a $N_2$-filled glove box.

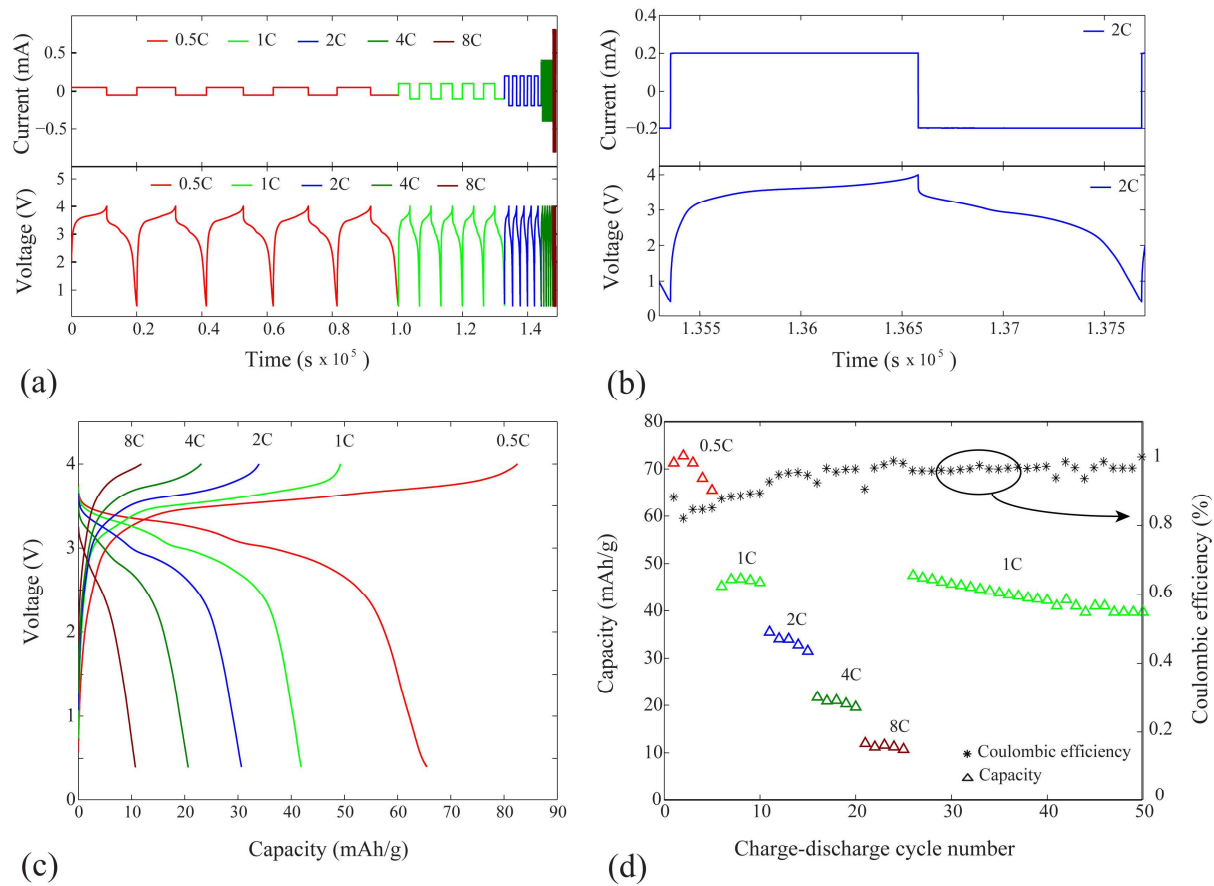

**Figure 4**. Electrochemical performance of the battery using tin@SPCT anode. (a) Cell voltages and currents in the first 25 cyclic charge-discharge tests with different C-rates. (b) A zoomed view of the cell voltage and current in a charge-discharge test with a 2 C rate. (c) Battery voltages in different charge-discharge cycles as a function of the battery capacity. (d) Specific capacity and coulombic efficiency of the battery measured in 50 charge-discharge cycles with different C rates.

The cyclic charge-discharge tests of this full fiber-shaped LIB using the tin@SPCT anode and LFP@SPCT cathode were performed with different charge-discharge rates (from 0.5 C to 8 C) as shown in Fig. 4(a, b). The battery open-circuit voltage was ~ 3.2 V. As shown in Fig. 4 (c, d), the battery had specific capacities of 67.1, 47.3, 35, 20.4, 12.5 mAh/g at 0.5, 1, 2, 4, and 8 C current rates, respectively. The measured specific capacities for a battery with a tin@SPCT anode are on average 10-30% higher than those of the battery using LTO@SPCT anode. Finally, we have returned to 1 C rate and performed 25 additional charge-discharge cycles (Fig. 4(d)) and found that specific capacity of the battery was still over 86% of the original. The coulombic efficiency of the battery was above 81% during all the charge-discharge tests, while mostly staying above ~93%. This indicates that the battery internal electrochemical structure remained intact even after been subjected to high current rates.

EIS studies of the ~12 cm long battery showed that the $R_s$ and $R_{ct}$ of the battery are ~134 Ω and ~276 Ω, respectively, according to the EIS spectrum as shown in Fig. 5(a), which are significantly smaller than the corresponding resistances of the LIB with LTO@SPCT anode reported in the previous section. This could be partially due to the fact that the measured resistance of the battery actually include the resistance of the electrodes, bulk electrolyte resistance and interface resistance between the electrolyte and the electrodes. Apparently, the electrical resistance of Tin@SPCT electrode is much smaller than that of the LTO@SPCT. Besides, the surface condition and thickness of the coated electrolyte layer and electrode layer may vary for the individual batteries, which may also responsible the difference in the battery resistances. In fact, the interfacial resistance $R_{ct}$ could be influenced by many factors including lattice mismatch, existence of the $Li^+$ deficient space charge regions in the solid-state electrolyte, the formation of interphases on the electrodes, and compatibility of the electrode material with the solid-state electrolyte [19]. A detailed comparison of $R_{ct}$ between the two batteries is out of the scoop of this paper.

Finally, we performed battery bending test using the same setup as described in the previous section. The total of 31000 bend-release movements were performed during 20 charge-discharge cycles of the battery. The battery specific capacity and coulombic efficiency during bending test are shown in Fig. 5(b). While the battery capacity decreased by 14% after 20 charge-discharge cycles, the coulombic efficiency of the battery remained greater than 84% during the whole test, while mostly staying above ~90%.

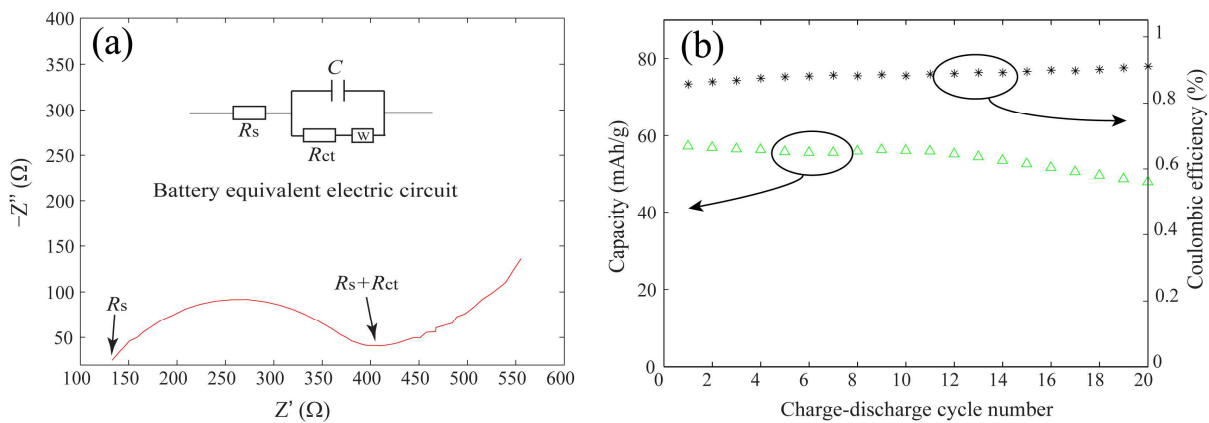

**Figure 5**. (a) EIS spectrum of the battery using a tin@SPCT anode. Insert: the equivalent electric circuit of the battery. (b) Battery specific capacity and coulombic efficiency of the battery during bending test.

**(3) LIBs comprising a LFP@SPCT cathode and a tin wire anode**

In this section we report the fiber-shaped battery comprising a LFP@SPCT cathode and a tin wire anode in order to compare its performance to a tin-covered SPCT anode. Fabrication of this battery follows the same procedures as detailed in the previous section with the only modification that we have replaced the tin@SPCT anode by a bare tin wire with a diameter of 500 μm.

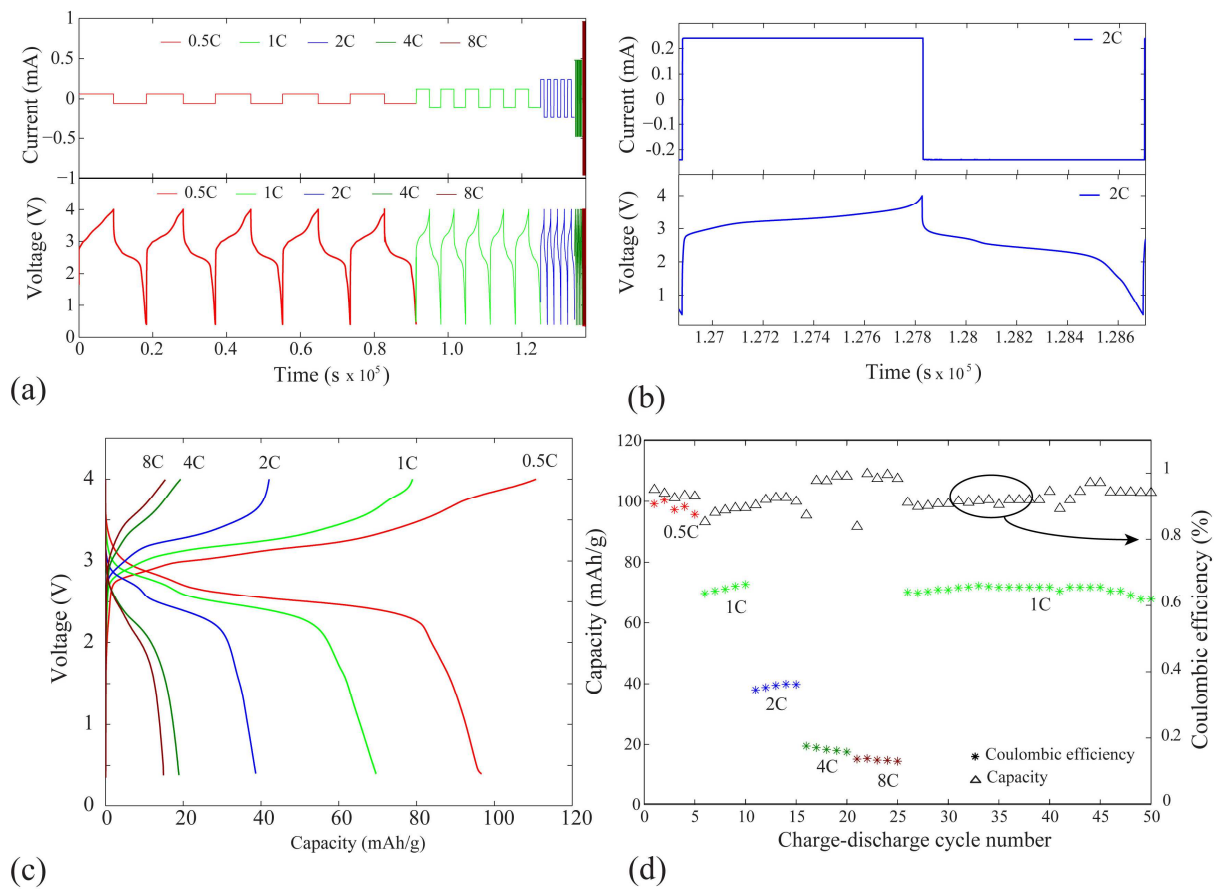

**Figure 6**. Electrochemical performance of the battery using a tin wire anode. (a) Cell voltages and currents in the first 25 cyclic charge-discharge cycles with different C-rates. (b) An enlarged view of the cell voltage and current in a charge-discharge test with a 2 C rate. (c) Battery voltages in different charge-discharge cycles as a function of the battery capacity. (d) Specific capacity and coulombic efficiency of the battery measured in 50 charge-discharge cycles with different C rates.

In Fig. 6, we present results of the battery charge-discharge tests at different C rates. We found the battery specific capacities to be 95.7, 72.3, 40, 19.4, 17.5 mAh/g at 0.5, 1, 2, 4, and 8 C current rates, respectively, which are the highest capacities of all the batteries presented in this work. The measured specific capacities for a battery with a tin wire anode are on average 10-30% higher than those of a battery using tin@SPCT anode. After a total of 50 charge-discharge cycles at different C rates, the battery specific capacity decreased only by 7.5%. The coulombic efficiency of the battery remained above 83% during the whole tests, while staying mostly above 91%. EIS measurements show that the $R_s$ and $R_{ct}$ resistances of the 12 cm long battery are ~97 Ω and ~190 Ω, respectively (see Fig. 7(a)), which are the lowest values among all the batteries presented in this work, while still comparable to those of a battery using tin@SPCT anode. Finally, during bending test (see Fig. 7 (b)), the coulombic efficiency of the battery with tin wire anode was always above 85%, while the battery specific capacity dropped only by 8%. As before, the test included 20 charge-discharge cycles with 1 C current rate during which the battery was subjected to 31000 bend-release motions.

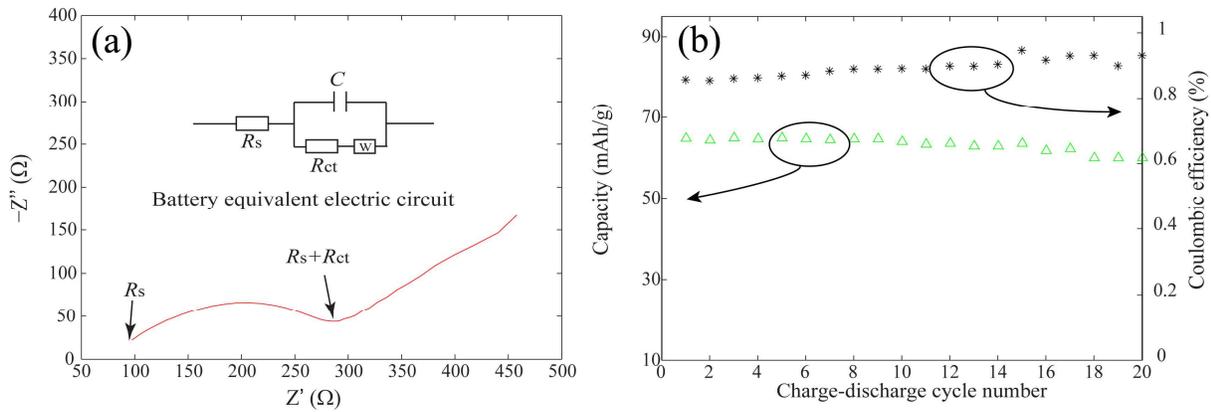

**Figure 7.** EIS spectrum of the battery using a tin wire anode. Insert: the equivalent electric circuit of the battery. (b) Battery specific capacity and coulombic efficiency during bending test.

**(4)   LIBs comprising a LFP@SPCT cathode and a SPCT anode**

For comparison, we also fabricate a LIB using a LFP@SPCT as cathode and a SPCT as anode. The battery fabrication process was the similar with that of the battery using the tin wire anode, except that the tin wire is simply replaced by a SPCT. The charge-discharge tests were performed to characterize this type of batteries. However, we found that batteries using the SPCT anode always showed a rapid degradation in their performance. A typical charge-discharge test of the battery using the pure SPCT anode is shown in Fig. 8. The coulombic efficiency of the battery was typically smaller than 35% and kept decreasing with more charge-discharge cycles. We, therefore, conclude that SPCT by itself is not an appropriate anode material.

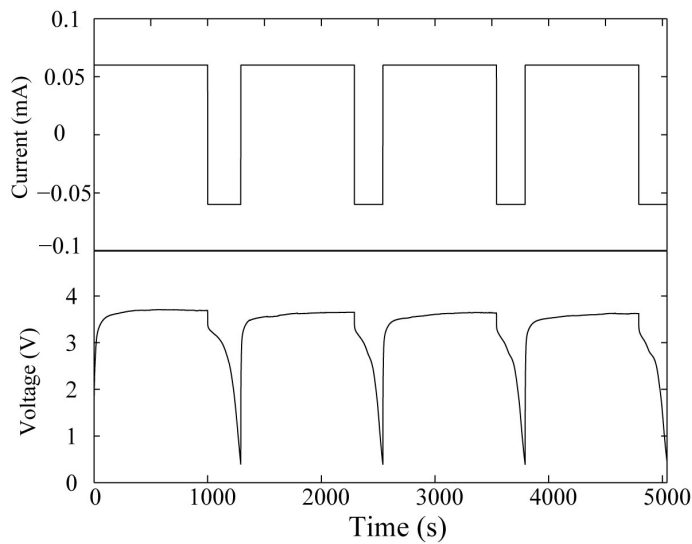

**Figure. 8**. A typical charge-discharge test result of a LIB using a pure SPCT anode.

**Conclusions**

In summary, we reported fabrication of all-solid, fiber-shaped LIBs assembled by co-twisting a cathode filament with an anode filament. The cathode filament was fabricated by depositing a LFP composite layer on a steel-filled polyester conductive thread (SPCT) via a dip-and-dry process. As anode filaments, we explored several material combinations including the LTO-composite coated SPCT (dip-and-dry deposition), tin-coated SPCT (PVD deposition) as well as a tin wire. The electrolyte composite layer consisting of $LiPF_6$ and PEO was then deposited onto both the anode and cathode filament before the battery assembly. By co-twisting the cathode filament and anode filament together using a customized jig, the batteries were thus assembled in a $N_2$-filled glove box. The open-circuit voltage was found to be ~ 2.3 V for the battery using the LTO@SPCT anode, and ~3.3 V for the batteries using the tin@SPCT anode and the tin wire anode. Charge-discharge tests were carried out at different C rates for each of the battery types. Experimental results show that the LIBs using the LTO@SPCT anode, the tin@SPCT anode and the bare tin wire anode achieve their maximal specific capacities of ~64, ~67, and ~96 mAh/g, respectively, when operated at 0.5-C rate. All the battery types showed significant, however reversible, reduction in their specific capacities when operated with higher C rates (Peukert constant of ~2). That said, the batteries retained well their specific capacities after 80 of the charge-discharge cycles, while showing Coulombic efficiencies greater than 90% most of the time. Moreover, even after been subjected to tens of thousands of bend-release motions, the batteries still retain their electrochemical performance and show only modest (less than ~10%) decrease in their specific capacities and Coulombic efficiencies. Overall, the fabrication of the proposed LIBs is simple and cost-effective, while the battery component materials are readily available commercially. Moreover, the detailed LIBs feature all-solid electrodes and electrolytes, thus avoiding safety concerns associated with electrolyte leakage, which makes them particularly suitable for wearable applications.

**Experimental Section**

(1) **Fabrication of the electrode filaments**

To fabricate the LFP@SPCT cathode filaments, a polyvinylidene fluoride (PVDF) solution was first prepared by dispersing 1g PVDF (powder, Mw. ~534,000, Sigma-Aldrich) into 10 ml 1-Methyl-2-pyrrolidinone (NMP) solvent (99.5%, Sigma-Aldrich) using a magnetic stirrer for 2 hours at room temperature (22℃). 0.425 g LFP (Phostech Lithium Inc.) was manually ground for 10 minutes and then mixed with 0.025 g carbon nanofibers (Sigma-Aldrich). The mixture was then dispersed into 2 ml as-prepared PVDF solution using the magnetic stirrer for 6 hours at room temperature (22℃). The SPCTs were rinsed by water and isopropanol for 15 minutes each in the ultrasonic bath. The cathode filaments were then fabricated by depositing a LFP composite layer onto the SPCTs via a dip-and-dry process performed in the $N_2$-filled glove box.

Fabrication of the LTO@SPCT anode filaments follows a similar route as detailed above, except that the LFP was replaced by LTO (Sud-Chemie Inc.) of the same weight.

The tin@SPCT anode filaments were produced by depositing a ~1.6 μm-thick tin (99.9%, Sigma-Aldrich) layer on the rinsed SPCTs by PVD (evaporative deposition, Edwards High Vacuum Ltd.). The PVD was performed under high vacuum (~$10^{-7}$ mbar) with a deposition rate of 0.2 nm/s. Then, the tin@SPCT wires were stored in the $N_2$-filled glove box.

The bare tin wire anodes (diameter: 0.5 mm, 99%, Sigma-Aldrich) were rinsed with regular detergent, 2 wt% HCl solution, and isopropanol for 20 minutes each in an ultrasonic bath. After rinsing, the tin wires were dried under the $N_2$ flow, and then stored in a $N_2$-filled glove box.

**(2) Preparation of the electrolyte solution, and deposition of electrolyte layer onto the electrode filaments**

0.125 g $LiPF_6$ (powder, Sigma-Aldrich) and 0.08 g $TiO_2$ (nanopowder, Sigma-Aldrich) was dispersed into 12.5 ml acetonitrile (99.9%, Sigma-Aldrich) solvent for 3 hours using a magnetic stirrer. Then, 0.665 g polyethylene oxide (powder, Mw ~400,000, Sigma-Aldrich) was added into the solution which was then stirred for 12 hours at room temperature (22℃). The as-prepared solution was cast into films onto both cathode and anode filaments to form an electrolyte layer which also functions as a battery separator. All the processes were carried out in the $N_2$-filled glove box.

**(3) Assembly of the fiber-shaped batteries**

The cathode and anode filaments were co-twisted into a battery using a home-made jig fabricated with a Makerbot 3D printer. The twisted LIB was then wetted by several drops of propylene carbonate. Finally, the LIBs were encapsulated within a heat-shrinking tube by heating them at 120℃ for 60 seconds. Both ends of the LIB were then sealed with epoxy.

**(4) Battery conditioning**

Before performing the charge-discharge tests of each battery, we need to carry out a battery conditioning process which would allow the formation of the nanostructures on the electrode that would withstand (or partially withstand) the volume change caused by the lithiation-delithiation process during the battery operation [20]. In the conditioning process, we generally used a 0.5 C current (calculated from the theoretical capacity of $LiFePO_4$) to charge-discharge the battery with different charging periods. In particular, we started with a charging period of 1 min, and charge-discharged the battery for 4 cycles. Then, we gradually increased the charging period from 1 min to 30 min with a step of 5 min, and for each charging period we run 4 charge-discharge cycles. During the conditioning process, we could also see that the coulombic efficiency of the battery increased gradually from ~60% to higher than 80%.